# Role of structure and charge trapping on the bipolaron formation and magnetic-field response of gated conjugated polymers

Zuchong Yang,[1] Vincent Lemaur,[2] Melissa Berteau-Rainville,[1] Olivier Bardagot,[3] Yoann Olivier,[4] Emanuele Orgiu[1]

[1] Centre Énergie Matériaux Télécommunications, Institut national de la recherche scientifique (INRS), 1650 boulevard Lionel-Boulet, Varennes, Québec, J3X 1P7, Canada

[2] Laboratory for Chemistry of Novel Materials, University of Mons, 7000 Mons, Belgium

[3] Institute of Chemistry and Processes for Energy Environment and Health (ICPEES), University of Strasbourg, CNRS, 25 rue Becquerel, Strasbourg 67087, France

[4] Laboratory for Computational Modeling of Functional Materials, Namur Institute of Structured Matter, Université de Namur, Namur B-5000, Belgium

Corresponding email: yoann.olivier@unamur.be , emanuele.orgiu@inrs.ca.

## Abstract

Conjugated polymers exhibit unique spin-dependent phenomena arising from weak yet critical hyperfine interactions. Understanding these spin effects, particularly the spin-dependent formation and decay of correlated spin pairs, is important for advancing both organic electronics and polymer-based spintronics. Intrinsic magnetic-field responses such as magnetoresistance have primarily been investigated in diode architectures, where electrons and holes coexist. However, such systems are less suitable for probing bipolaron formation in unipolar transport, and the relationship between polymer structure and bipolaron formation in lightly doped polymers remains unclear. Here, we systematically investigate intrinsic magnetoresistance in representative conjugated polymers using field-effect transistors and observe a generally positive magnetoresistance. First-principles simulations reveal that bipolarons preferentially form on short conjugated segments associated with amorphous regions. Moreover, comparisons across these polymers show that enhanced charge trapping correlates with stronger magnetoresistance, implying promoted bipolaron formation. Bipolaron-incorporated energy-level-alignment modeling near metal/polymer interfaces suggests that charge traps can increase the bipolaron density.

Certain migratory animals possess a magnetic compass sense of the Earth's magnetic field for navigation, which was mainly ascribed to the spin-dependent chemical reactions of organic radical pairs[1,2]. The latter are also responsible for magnetic control of spin dynamics in fluorescent proteins[3,4]. Likewise, in π-conjugated polymers mainly composed of light elements with weak spin-orbit coupling and having a relatively low degree of structural order, electrical conductivity and luminescence also display magnetic response. The most notable example is the magnetoresistance observed in organic light-emitting diodes with non-ferromagnetic contacts[5]. To date, little is known about how conjugated polymer structure is related to their intrinsic response to magnetic fields. More knowledge must be garnered to improve the design and efficiency of organic electronic devices. Furthermore, a deeper understanding of spin-dependent properties in such polymers can contribute to the advancement of spintronics, where electron spin is used for information storage and processing.

The most commonly proposed mechanisms to rationalize organic magnetoresistance are the magnetic-field effects (MFEs) on the singlet-triplet mixing of a pair of charge carriers. These mechanisms differ in compositions of the spin pair and how the corresponding spin-dependent pair formation and decay affect charge transport[6]. Distinct mechanisms could dominate depending on material composition, electrode configuration, light irradiation, or applied voltages[7-9]. The mechanistic complexity is even more challenging to disentangle if one takes into account the injection/transport of both electrons and holes as well as the coexistence of excitons and unbound charges in the conventionally used diode architecture. This also poses a challenge to examining the bipolaron mechanism proposed in a truly unipolar transport scenario[10-12]. Additionally, during the typical diode fabrication process, the possible penetration of metal from the top electrode into the underlying organic layer might lead to the formation of undesired metal-organic complexes or shunt resistance, which could distort the proper interpretation of magnetoresistance results. Within this context, field-effect transistors (FETs) represent a suitable platform as the type and concentration of charge carriers in the channel are, in principle set by the intensity of the gate field, and metal-penetration issues can be conveniently alleviated by depositing the active materials on top of pre-deposited electrodes (bottom-contact configuration). Moreover, working with planar FETs also enables probing the role of MFE in state-of-the-art polymers, generally designed for conducting in the x-y plane.

MFE on bipolaron formation is central to explaining unipolar magnetoresistance[12,13]. The bipolaron is a quasiparticle consisting of a pair of likely charged polarons in a singlet state, i.e., two holes or two electrons sharing the same deformed molecular segment (intrachain) or two π-conjugated segments (interchain). Due to Coulombic repulsion, there will be a significant energy cost, Hubbard penalty ($U$), for its formation. In highly doped polymer systems via molecular or electrochemical doping, the bipolaron spectral features have been reported recently[14-17]. In contrast, whether they can be formed in cases through bare electrical injection or electrostatic gating is far less explored and more challenging to be characterized by techniques such as absorbance spectroscopy or electron paramagnetic resonance. Studying unipolar magnetoresistance in FETs thus offers an alternative pathway to investigate this topic. Furthermore, a significant energetic disorder present in materials or at the device interface was usually assumed to account for $U$ to rationalize the existence of bipolarons to interpret magnetoresistance[18,19]. Hitherto, a clear relationship that links polymer structure to bipolaron formation is still lacking, despite its critical significance to fully leverage the advantage of polymers in terms of tailorable chemical structure.

Here, we investigated bipolaron formation in several representative conjugated polymers using the FET architecture by comprehensively tuning temperature, electric field intensity, and device interface. In general, we found a positive magnetoresistance related to bipolaron formation at cryogenic temperatures. Density function theory (DFT) calculations were employed to study the intrinsic structural footprint affecting the probability of bipolaron formation. Bipolarons in singlet configuration were found to be more stable than polaron pairs in triplet configuration only on short conjugated segments. The latter was found to prevail in the amorphous phase of polymers by molecular dynamics (MD) simulations. Moreover, by comparing the magnetoresistance of different polymer structures and chemically-functionalized interfaces, we further established a correlation between the magnetoresistance amplitude and charge trapping time constant characteristic of the devices. An expanded energy-level-alignment (ELA) modeling framework incorporating both bipolarons and traps suggests that charge trapping can boost the bipolaron density at the metal/polymer interface.

**Positive magnetoresistance at cryogenic temperatures**

We measured magnetoresistance in a bottom-gate bottom-contact transistor, while an external magnetic field ($B$) was applied perpendicular to the substrate under high vacuum (Fig. 1a). Three representative conjugated polymers were investigated (Fig. 1b), including $p$-type regioregular poly(3-hexylthiophene-2,5-diyl) (rr-P3HT), $n$-type poly(*N,N′*-bis-2-octyldodecyl-naphtalene-1,4,5,8-bis-dicarboximide-2,6-diyl-*alt*-2,2′-bithiophene-5,5′-diyl) (P(NDI2OD-T2)), and ambipolar bithiophene-isoindigo donor-acceptor copolymer with branched side chains positioned at the fourth carbon atom (PII2T-C4).

The magnetoresistance was first measured in three polymer-FETs biased in the corresponding $p$- or $n$-type unipolar saturation regime (Fig. 1c–e). In general, a positive, Lorentzian-like saturation line shape could only be observed at temperatures below 100 K. Interestingly, PII2T-C4 has a larger magnetoresistance linewidth and amplitude than the other two polymers. The saturation persists up to 500 mT (Fig. 1f), which is distinct from the others, where a negative high-field ($B$ > ~250 mT) component would emerge (Supplementary Fig. 3a). These line shapes can be fitted by the "non-Lorentzian" model (equation (1)), where $MR_{sat}$ and $B_0$ quantify the saturation amplitude and linewidth, respectively. The temperature-dependent $MR_{sat}$ and $B_0$ are shown in Fig. 1g, h. $MR_{sat}$ of PII2T-C4 biased with $p$- or $n$-type saturation condition (Supplementary Fig. 3b) is relatively insensitive to temperature up to 75 K, while that of P3HT and P(NDI2OD-T2) gradually diminishes to zero. Moreover, the $p$-type $MR_{sat}$ is larger than the $n$-type one for the same PII2T-C4 device. The $B_0$ values of P3HT and P(NDI2OD-T2) are on average 0.9 and 1.3 mT, respectively, which are smaller than 8.5 (measured with 100 mT range) or 14.9 mT (500 mT range) for PII2T-C4.

$$MR = MR_{sat} \frac{B^2}{(|B| + B_0)^2} \tag{1}$$

The low-field magnetoresistance originates from the competition between $B$ and the intrinsic hyperfine field ($B_{\mathrm{hf}}$)[20]. The latter describes the strength of hyperfine interactions between spin and nuclear magnetic moment of nearby atoms. It is usually no larger than 3 mT[21], hence the magnetoresistance is typically found to be prominent up to tens of mT and quickly enters saturation afterwards. $B_0$ is in principle positively correlated with $B_{\mathrm{hf}}$. Nevertheless, considering the similar chemical compositions of donor and acceptor units in PII2T-C4 and P(NDI2OD-T2), it is unlikely that their $B_{\mathrm{hf}}$ differ as significantly as the measured $B_0$.

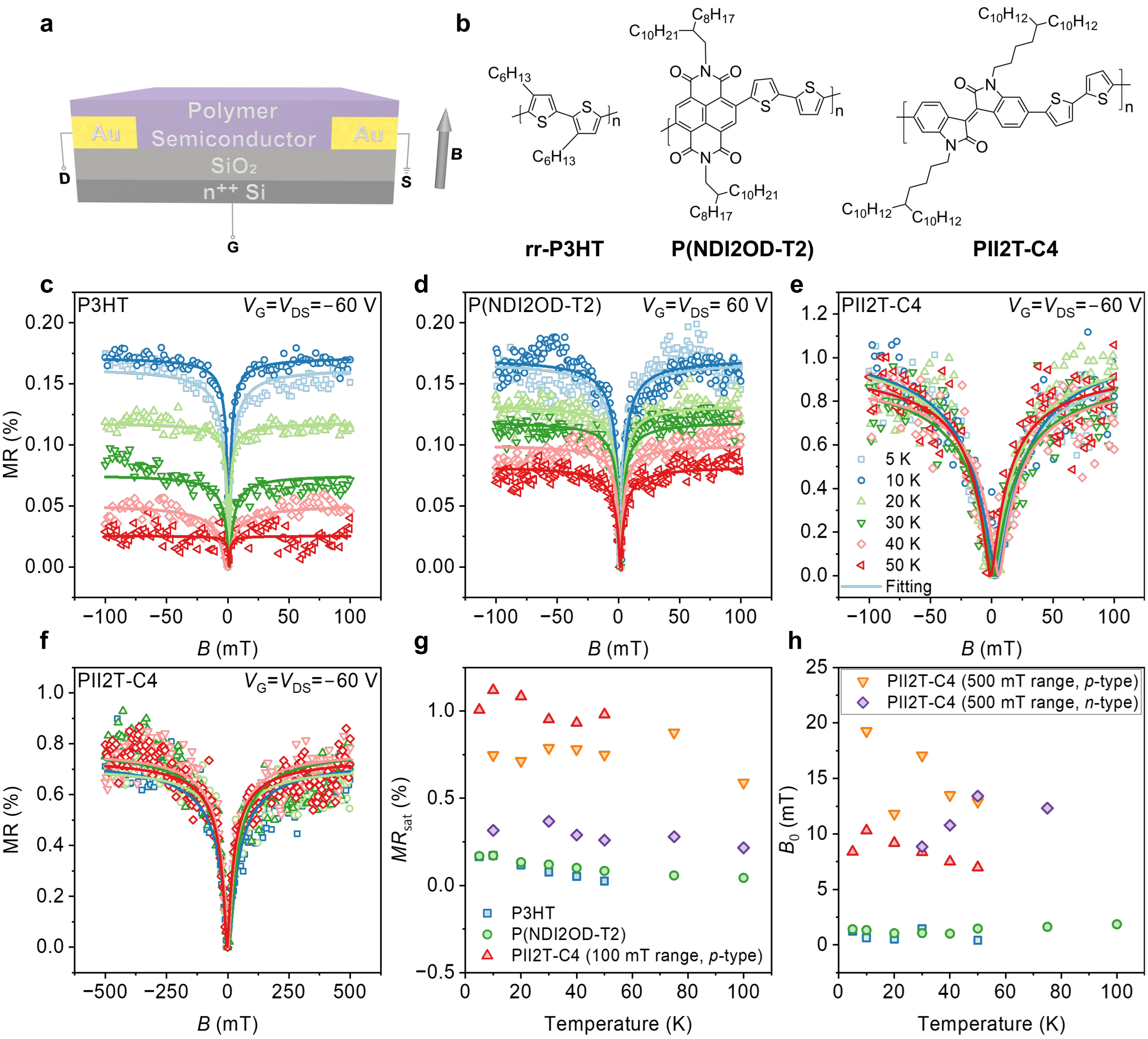

**Fig. 1 | Positive magnetoresistance in conjugated polymers-based FETs below 100 K.** a) Schematic of device structure. b) Chemical structures of three polymers. c)–e) Magnetoresistance line shapes with $B$ scanned in the range of ± 100 mT for P3HT, P(NDI2OD-T2), and PII2T-C4, respectively. The line shapes are reproducible in two sets of measurements. f) Magnetoresistance line shapes of PII2T-C4 measured with another device using $B$ range of ± 500 mT. g), h) Temperature-dependent $MR_{sat}$ and $B_0$ by fitting the corresponding line shapes with the non-Lorentzian model, respectively. The results of PII2T-C4 under *p*-type ($V_G$ = $V_{DS}$ = –60 V) or *n*-type ($V_G$ = $V_{DS}$ = 90 V) saturation conditions measured with the same device and 500 mT range are shown in both figures.

## Bipolaron formation and the impact of charge trapping

According to the bipolaron magnetoresistance mechanism[12], when two neighbouring spins naturally precess around their respective local $B_{hf}$, which have random orientations, a bipolaron can form in a singlet configuration as allowed by the Pauli exclusion principle. Instead, when $B$ is applied ($B \gg B_{hf}$), the spins of two polarons can only precess about $B$, which will force their spins to be parallel. Hence, they will possess a triplet character, preventing bipolaron formation, which is called the spin-blocking effect. Here, a presumption is that the bipolaron formation is relevant in the charge hopping/percolation process[22,23], that is, a polaron would have a finite probability to hop to a nearby site that has been occupied by another polaron rather than directly hopping to other empty sites. The ratio of these two hopping probabilities was called the branching ratio[12]. A localized hopping range and limited hopping dimensionality could favor bipolaron formation which ultimately affects charge transport. This model implies that the higher the branching ratio, the larger the $B_0$[12]. Thus, a straightforward hypothesis is that bipolaron formation is more probable in the PII2T-C4 device. To validate this hypothesis, we first analyzed their charge transport properties. The transfer characteristics (Supplementary Fig. 4) reveal that, above 125 K, all of them show hysteresis indicative of charge trapping, with the most significant in PII2T-C4. Importantly, below 100 K (Fig. 2a–c), unlike P3HT and P(NDI2OD-T2), PII2T-C4 still displays a large and counterclockwise hysteresis (Fig. 2h).

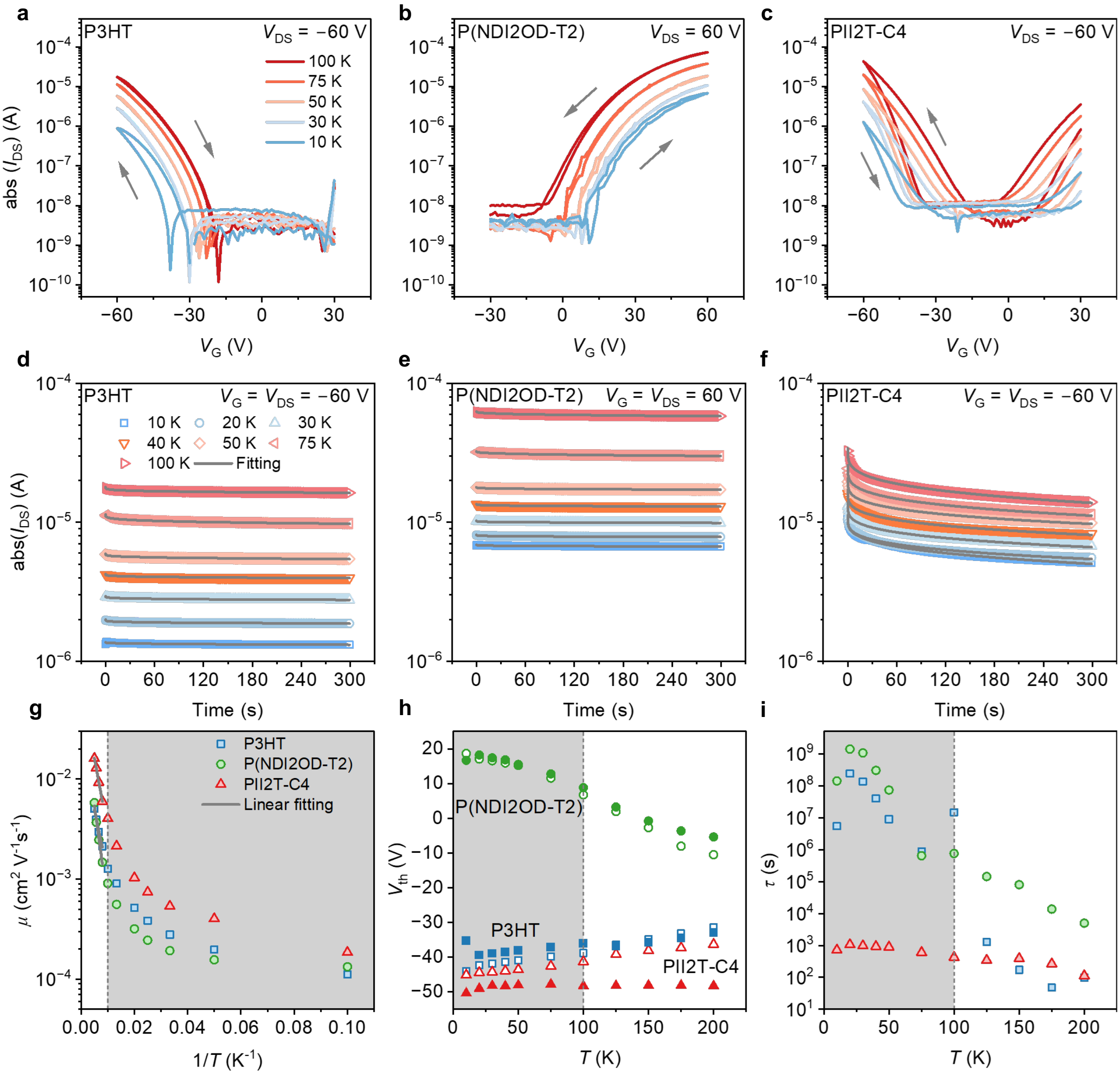

**Fig. 2 | Low-temperature charge transport and trapping**. a)–c) Saturation transfer characteristics at various temperatures below 100 K. d)–f) Bias stress-induced current decay with time. The decay can be well fitted by the stretched exponential model. g) Arrhenius plot of temperature-dependent saturation mobilities. h) Temperature-dependent threshold voltages in the forward (open symbols) and reverse (closed symbols) scans. i) Temperature-dependent charge trapping time constant extracted from the stretched exponential model. All the shaded areas indicate the below-100 K regime.

Another distinct behavior of the PII2T-C4 device is the more significant bias-stress effect (Fig. 2d–f). The current decay with time can be fitted with a stretched exponential:

$$I_{\mathrm{DS}}(t) = I_{\mathrm{DS}}(0)\exp\left[-\left(\frac{t}{\tau}\right)^{\beta}\right] \tag{2}$$

where $\tau$ is a time constant that is characteristic of the mean time for charges staying mobile before encountering traps, while $\beta$ $(0 \le \beta \le 1)$ is a dispersion exponent related to the distribution of trapping energy barriers[24,25]. While all polymers show a comparable $\beta$ below 100 K (Supplementary Fig. 5), $\tau$ in PII2T-C4 is at least three orders of magnitude lower (Fig. 2i), which reveals a more significant charge trapping. Since the bipolaron formation mechanism requires an already occupied site, charge trapping can facilitate bipolaron formation by compensating for the energy cost $U$[26], thus making bipolaron formation more likely in PII2T-C4.

The temperature-dependent mobilities of all polymers (Fig. 2g) exhibit a typical thermally activated hopping behavior above 125 K, with PII2T-C4 showing the highest mobility. Their film microstructures were characterized by grazing incidence wide angle X-ray scattering (GIWAXS). As shown in Supplementary Figs. 9 and 10, PII2T-C4 displays well-resolved high-order lamellar packing. PII2T-C4 has the lowest paracrystalline disorder in the π–π stacking direction (Supplementary Table 1), which aligns with its highest mobility. Low-temperature charge transport measurements reveals a weaker temperature dependence than the high-temperature one, with hopping becoming more localized and dimensionally limited in this regime[27-29]. This regime is thus suitable to examine the bipolaron mechanism. Meanwhile, this may explain why there was no positive magnetoresistance measurable above 125 K, unlike the previous reports in vertical devices. Since field-effect mobility is usually several orders of magnitude higher than in diodes for the same material[30], the presence of more delocalized charges at high temperatures would go against bipolaron formation or its relevance for charge transport.

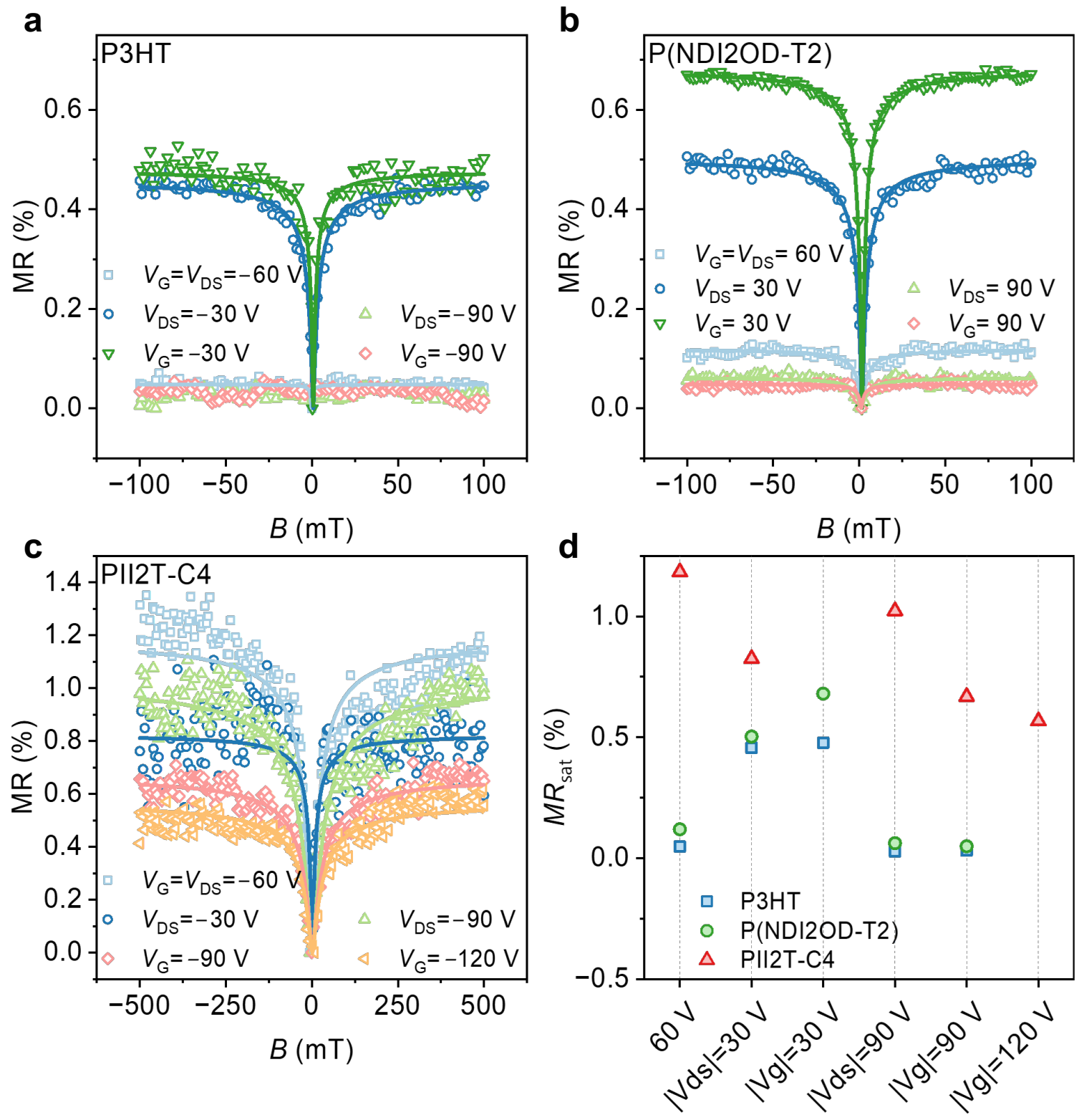

**Fig. 3 | Electric-field effects on the positive magnetoresistance**. a)–c) Magnetoresistance line shapes at various biasing conditions of P3HT, P(NDI2OD-T2), and PII2T-C4, respectively. Compared to the reference measurements with the same $V_G$ and $V_{DS}$, the $V_{DS}$ ($V_G$) was varied, with $V_G$ ($V_{DS}$) kept constant. d) $MR_{sat}$ as a function of voltages in three polymers.

According to the electric field-assisted hopping mechanism, enhancing electric fields increases the hopping probability at low temperatures[28]. Thus, one would expect reduced magnetoresistance upon higher electric field, which is indeed experimentally observed (Fig. 3a–c). As summarized in Fig. 3d, three polymers consistently show a lower $MR_{sat}$ at larger $|V_G|$, although it is not as sensitive to $|V_{DS}|$ in PII2T-C4. Interestingly, Wang et al. recently reported a decreased electrically detected magnetic resonance signal under high electric fields in polymer FETs. High electric fields enabled more percolation pathways surpassing those contain traps necessary for bipolaron formation[31]. The consistent electric-field dependence of two distinct macroscopic observables related to spin pair dynamics points out the critical role of bipolarons in a unipolar conduction system.

## Intrinsic structural effect on bipolaron formation

In electrochemical doping of polymers, the amorphous region of polymers was suggested to be the key to accommodate bipolarons[16,32]. Spectro-electrochemical measurements were employed to characterize the bipolaron signatures in these polymers under electrochemical doping. As shown in Supplementary Figs. 15–18, the bipolaron spectral features for each polymer were observed except for the $n$-doping case (i.e., negative bipolarons) in PII2T-C4 (see Supplementary Section 4). The results demonstrate the complexity of bipolaron formation in the electrolyte-gated case, which is also related to other extrinsic factors such as the solvent swelling capability[33] and counterion location[34].

Unlike the electrochemical doping case, where the bipolaron could be coulombically stabilized by oppositely charged counterions, bipolarons in cases through electrical injection or electrostatic gating should be more scarce and might exist as an intermediate/transient state formed along the hopping pathway[35]. To unveil the intrinsic structural factor related to bipolaron formation in these cases, we first carried out DFT calculations to evaluate the stability of bipolaron (in singlet configuration) with respect to polaron pair (triplet) as a function of chain size (number of units). Note that in P3HT, one unit corresponds to one thiophene, while in P(NDI2OD-T2) and PII2T-C4, one monomer was considered to consist of three and four units, respectively. As highlighted in Fig. 4a, bipolarons in P3HT are stable up to pentamers (18 Å), in line with the previous study[36], while those in P(NDI2OD-T2) and PII2T-C4 are found to be roughly localized on one monomer (13 and 17 Å, respectively). The spatial localization of doubly charged states depicts that the isoindigo of PII2T-C4 strongly localizes electrons (similar to the NDI in P(NDI2OD-T2)), which stabilizes the negative bipolaron on short segments, while the holes are more delocalized (Fig. 4b and Supplementary Fig. 19). On the monomer of PII2T-C4, the negative bipolaron is far more stable than the positive one. Indeed, the close proximity of the two charges with same spin orientation is unfavorable in virtue of Pauli's principle. The energy cost to convert the negative bipolaron to polaron pair (1.17 eV) is much higher than in the positive case (0.23 eV). Overall, the results show that bipolaron formation is not always intrinsically favored from the structural perspective, the predominance of bipolarons over polaron pairs preferentially occurs at short conjugation length.

Conjugated polymers typically have heterogeneous microstructures with π-conjugated aggregates separated by amorphous regions with higher conformational disorder[37], where the conjugation is more likely to break, and could be an ideal site to accommodate bipolarons. To look more closely into the conjugation in the amorphous region, MD simulations were run for all polymers (see Methods and Supplementary Fig. 20). We comprehensively analyzed the distribution of conjugation length for every individual chain, considering that the conjugation break occurs if the torsion angle between two adjacent segments is larger than the cutoff angle of 15, 30, or 45 degrees. From the distributions of all chains (Supplementary Fig. 21 and Supplementary Table 4), the median conjugation length is between 2 and 3 units. The most probable maximum conjugation length varies between 2 and 4 for P3HT, between 2 and 6 for P(NDI2OD-T2), and between 3 and 5 for PII2T-C4 depending on the cutoff angle, corresponding roughly to one monomer for the latter two polymers (Fig. 4c–e). These results reveal that in the amorphous region, the conjugation length is short, which falls within the suitable range to facilitate bipolaron formation as determined by DFT. The amorphous region is thus important for the bipolaron formation. Moreover, these simulations suggest that *p*-type magnetoresistance should be inherently higher than *n*-type one in PII2T-C4. Indeed, when $B$ is applied, two spins in the positive or negative doubly charged state localized on a monomer are aligned, so that the originally stable bipolarons will be forced into polaron pairs. The lower energy cost in the hole-hole case would enhance the MFE.

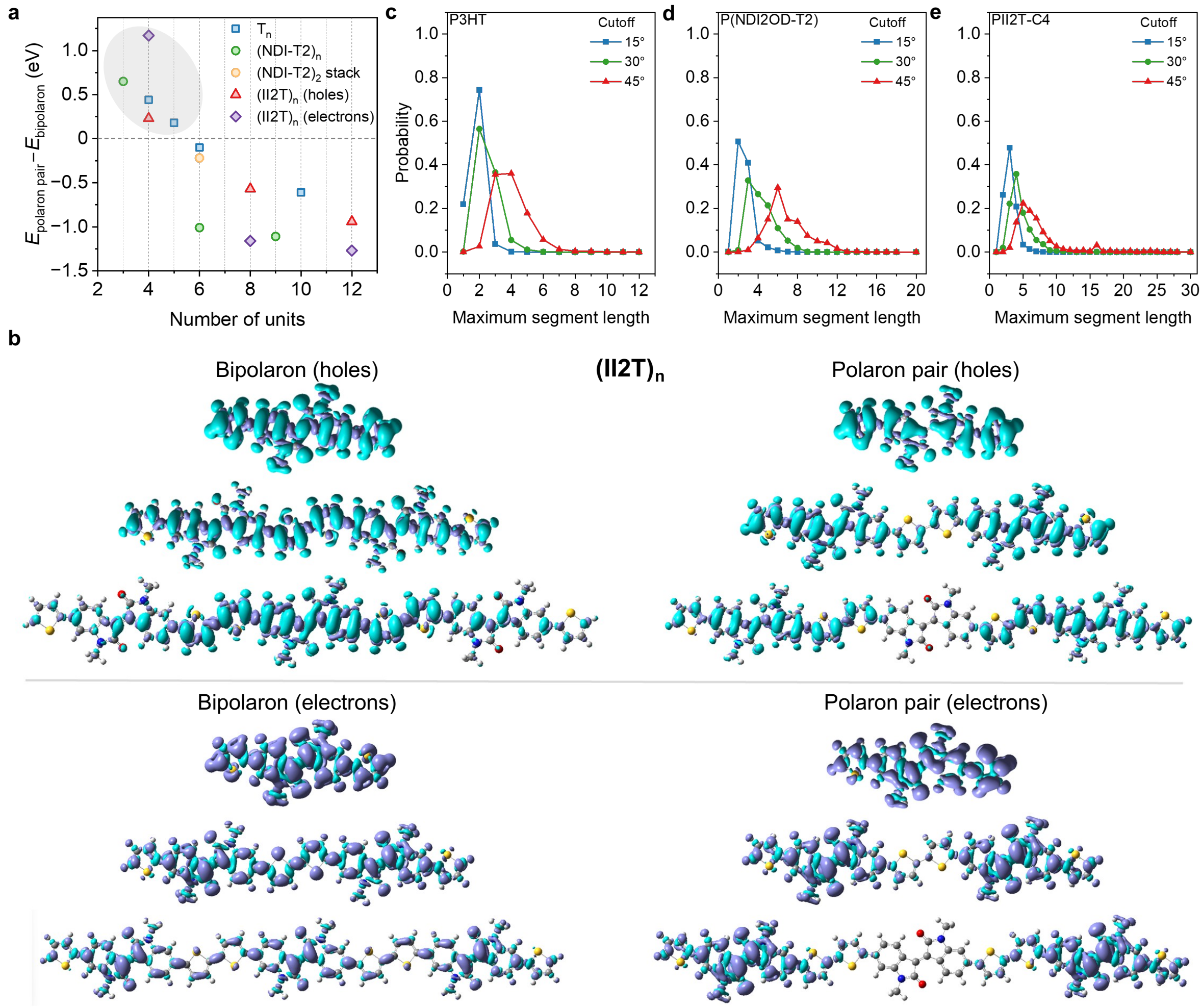

**Fig. 4 | Intrinsic structural effect on the formation of bipolaron by first-principles calculations.** a) DFT-computed energy difference between polaron pairs and bipolarons energies ($E_{polaron\,pair} - E_{bipolaron}$) as a function of chain size (number of units) using the oligomers corresponding to P3HT, P(NDI2OD-T2), and PII2T-C4. The case of two π-stacked two monomers for P(NDI2OD-T2) was also considered. For the ambipolar PII2T-C4, the doubly-charged states based on both two holes and two electrons were calculated. The shaded area highlights the situation where bipolarons are more energetically stable. b) Spatial localization of the positive/negative bipolarons and polaron pairs of $(II2T)_n$ as a function of chain length (n = 1, 2, 3). c)–e) MD-simulated probability distribution of the maximum segment length (number of units) under different twist-angle cutoffs in the amorphous phase of P3HT, P(NDI2OD-T2), and PII2T-C4, respectively.

## Negative magnetoresistance of PII2T-C4 at high temperature

Another peculiar behavior of PII2T-C4 is that it continues to display a negative MR above 125 K. As shown in Fig. 5a, this negative magnetoresistance has a less saturated line shape. The amplitude increases with temperature (Fig. 5b), and the linewidths become even larger than those at lower temperatures (Fig. 5c). The sign change was often ascribed to the shift of charge conduction type from unipolar to ambipolar, and hence a different mechanism based on, e.g., MFE on electron-hole pair recombination took over[38-40]. Specifically, the spin mixing between singlet and triplet states of the electron-hole pair is suppressed by *B*, which would yield fewer singlets. As the triplet in conventional organic semiconductors has a longer lifetime before recombination than the singlet, there would be less recombination, hence the device current is enhanced, i.e., yielding negative magnetoresistance[41].

This could be responsible for the abrupt change of magnetoresistance in the ambipolar PII2T-C4, including its sign, linewidth, and temperature-dependence. At elevated temperature, the thermionic field emission of minority electrons might be increasingly non-negligible at the drain electrode side, which has a net zero electric potential ($V_{\mathrm{GD}} = 0$ V). Such pinch-off region has a strong local electric field which would let electrons leak and then encounter holes (injected from the source electrode side) in the channel to form precursor electron-hole pairs before recombining. The recombination zone near the drain electrode under such biasing conditions for an ambipolar molecule was indeed observed previously[42]. Moreover, from the perspective of charge trapping, these electrons can also be regarded as a source of traps for the majority holes to be ejected. Therefore, the suppressed recombination by $B$ will weaken hole trapping and result in negative magnetoresistance.

We further measured the electric-field dependence in this regime (Supplementary Fig. 6). Enhancing the lateral electric field by either increasing $V_{\mathrm{DS}}$ or shrinking the channel would facilitate the electron injection and the encounter of electrons and holes, thus would produce a higher $MR_{sat}$, which aligns with the results (Fig. 5d). In contrast, $MR_{sat}$ has no apparent dependence on the gating field, which reveals that changing the hole concentration alone in the channel has no direct impact, different from the behaviour of low-temperature regime. Additionally, a pulsed-gating method (Supplementary Fig. 7) enables us to observe the negative magnetoresistance at room temperature (Fig. 5e).

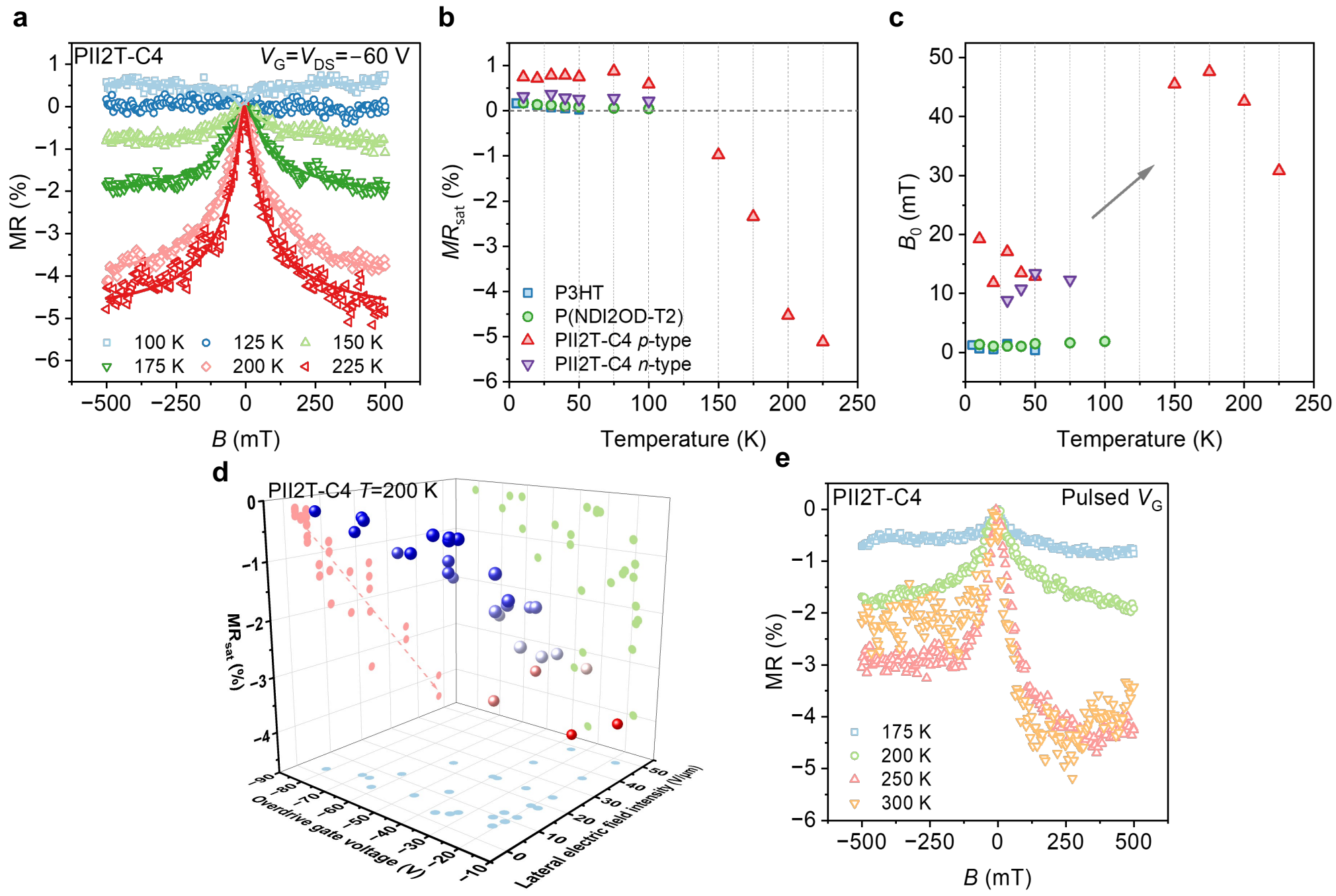

**Fig. 5 | Negative magnetoresistance above 125 K and electric-field effects in PII2T-C4.** a) Magnetoresistance line shapes of PII2T-C4 at temperatures above 100 K. b), c) Summary of $MR_{\mathrm{sat}}$ and $B_0$ at all temperatures, respectively. The abrupt changes of sign and linewidth are indicated. d) The effects of overdrive gate voltage and lateral electric field on $MR_{\mathrm{sat}}$ at a temperature of 200 K. The color scale from blue to red of the data points corresponds to the increasing amplitude. The dashed arrow in the $MR_{\mathrm{sat}}$ projected planes is a guide to the eye. e) Magnetoresistance measured with the pulsed-gating method from 175 K up to room temperature (300 K).

**Interface engineering and interfacial bipolaron modeling**

As charge transport in FETs is constrained in a limited volume with a thickness of a few nanometers close to the semiconductor/dielectric interface, it is sensitive to the interfacial profile. Thus, interface engineering can be a useful knob to tune the magnetoresistance. One approach is to modify the surface properties of dielectrics or electrodes with self-assembled monolayers (SAMs).

First, the dielectric surface was treated with various silane-based SAMs: hexamethyldisilazane (HMDS), trichloro(octadecyl)silane (OTS), perfluorooctyl-trichlorosilane (PFS). These silane-based SAMs were commonly employed to make the surface more hydrophobic by deactivating the dangling silanol groups[43]. The surface energy was confirmed by the contact angle measurements with the UV-ozone (UVO) treated sample as a reference (Supplementary Fig. 22 and Supplementary Table 5). The mobilities of these dielectrics-treated devices follow the order of UVO < PFS < HMDS < OTS (Supplementary Fig. 8a). The ambipolar transport is the most balanced in the OTS-treated device, reflected by the symmetric V-shaped curves. The electron transport in the former two devices was not observed, which indicates that trapping of electrons was more pronounced. The rich silanol groups and electron-withdrawing dipoles of the UVO- and PFS-treated surface respectively act as electron traps[44]. The microstructure of PII2T-C4 films deposited on these dielectrics show comparable edge-on-dominant crystalline textures (Supplementary Fig. 13). The reducing paracrystalline disorder from UVO- to OTS-treated samples well reflects the mobilities (Supplementary Table 3). Their magnetoresistance consistently shows a sign transition from positive to negative (Fig. 6a) and the order in $MR_{sat}$ correlates with $\tau$ (Fig. 6c)

Second, the electrode surface was functionalized with thiol-based SAMs, including 1-decanethiol (1DT) and perfluoro-decanethiol (PFDT). Our previous study has demonstrated that the work function of gold electrodes could be effectively tuned by these two SAMs with opposite dipole moments. Nevertheless, after contacting the polymer, the equilibrium Fermi level across the metal/polymer interface was pinned at ~ 0.79 eV above the highest occupied molecular orbital (HOMO) of PII2T-C4[45]. Even though the corresponding devices show a smaller difference in $\tau$ compared to the above dielectric case, their $MR_{sat}$ still correlates with $\tau$ (Fig. 6b, d).

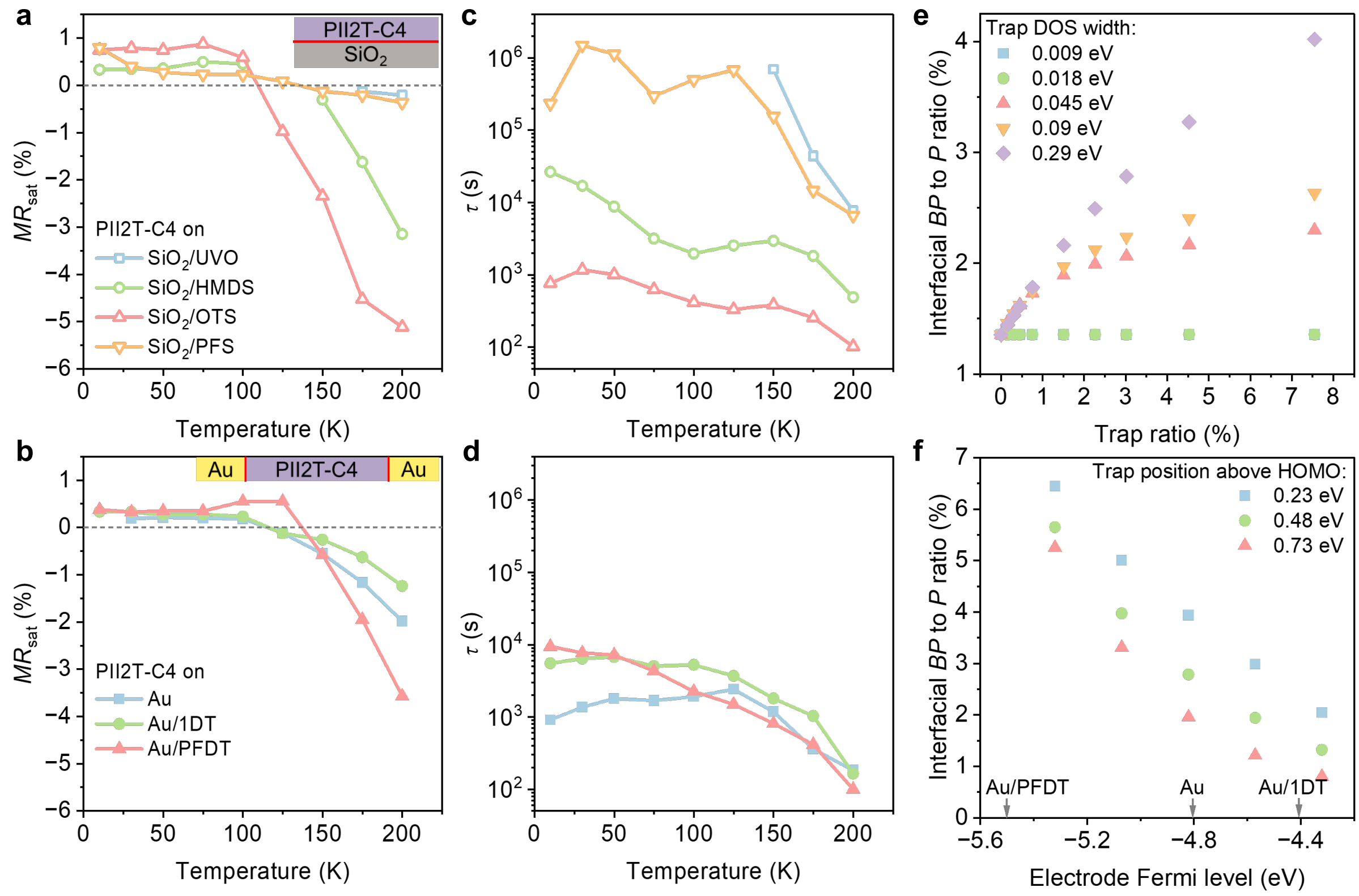

**Fig. 6 | Interface engineering of magnetoresistance and bipolaron modeling at the metal/polymer interface.** $MR_{sat}$ (a) and charge trapping time constant $\tau$ (c) for devices with different dielectric surface treatments. $MR_{sat}$ (b) and $\tau$ (d) for devices with different electrode functionalization. e) Interfacial bipolaron-to-polaron ratio as a function of designated trap ratio in polymers. The Hubbard energy was set at 0.3 eV, the HOMO density of states (DOS) width is 0.29 eV, and the electrode Fermi level is –4.82 eV for bare gold. f) Dependence of the bipolaron-to-polaron ratio on the Fermi level of the electrode at each trap position (above HOMO onset). The HOMO onset is –5.31 eV. The Fermi levels of bare gold, 1DT- and PFDT-functionalized gold are indicated.

It was proposed that the bipolaron concentration can be significant near metal/semiconductor contacts based on a modified energy level alignment (ELA) model[19]. There are several factors advantageous for the bipolaron formation at the interface than in the bulk, including potentially larger energetic disorder, Fermi-level pinning, and image-charge stabilization effect[46]. Here, our new ELA modeling framework allows for calculating the possible bipolaron concentration at the gold/polymer interface (see Supplementary Section 1.2). Notably, such an ELA framework also includes the density of states (DOS) of hole traps to simulate the trapping effect. As expected, the bipolaron concentration was found to be diminished with increasing energy cost $U$ (Supplementary Fig. 2b). In Fig. 6e, assuming $U$ of 0.3 eV, the results show that increasing the trap density facilitates the bipolaron formation if the trap DOS has enough width. Interestingly, the bipolaron concentration can be significantly boosted by deepening the electrode Fermi level (Fig. 6f), which aligns with the above measurements where the PFDT-functionalized device with the highest electrode work function[45] shows the largest magnetoresistance. The modeling results reveal that traps can indeed facilitate the bipolaron formation, particularly at the metal/polymer interface.

Lastly, regarding the reason for the higher trap density in the PII2T-C4 device compared to the others, we have excluded two possible sources: (i) the heavy-metal catalyst residues, using ICP-MS[47] and (ii) the intrinsically stronger proton affinity of polymer chains according to the protonation mechanism[48] (see Supplementary Section 8). Future studies may be able to unveil the origin of traps due to water-related protonation, for instance, using techniques such as four-dimensional scanning transmission electron microscopy[49] to compare the density of nanovoids which can accommodate water in these polymer films (see Supplementary Section 9).

## Conclusion

In conclusion, through systematically investigating the magnetoresistance using representative conjugated polymer-based FETs, we found generally positive magnetoresistance at cryogenic temperatures and a sign reversal only in PII2T-C4. The first-principles calculations demonstrate that bipolaron formation is structurally permitted even in the lightly doped case, which happens more probably in the amorphous region where the conjugation length is short enough. The structure-property relationship for bipolaron formation in polymers pinpoints the important role of the disordered domain together with the choice of electron-withdrawing and electron-donating units in donor-acceptor copolymers in understanding the unipolar magnetoresistance. From the device perspective, the bipolaron formation can be further facilitated by the charge trapping effect, especially at the metal/semiconductor interface. The interface engineering offers a useful tool to tune the charge trapping kinetics and magnetoresistance. Our work demonstrates that both intrinsic structural factors and charge trapping effects play critical roles in governing bipolaron formation and magnetoresistance in conjugated polymer-based devices. These findings pave the way for investigating bipolaron formation and magnetic-field response in other systems, such as highly doped conducting polymers and organic–inorganic perovskites.

## Methods

### Materials

rr-P3HT ($M_w$ > 45,000, regioregularity > 93%) was purchased from Luminescence Technology Corp., P(NDI2OD-T2) ($M_w$ = 50,000–100,000, Đ = 2–3), and PII2T-C4 ($M_w$ = 78,000, Đ = 2.4) were purchased from 1-Material Inc. The HMDS, OTS, PFS, 1DT, PFDT, *o*-DCB (anhydrous, > 99%), heptane (anhydrous, > 99%), and 2-propanol (anhydrous, > 99.5%) were purchased from Sigma. Chloroform (extra dry and Amylene-stabilized, > 99.9%) and Toluene (anhydrous, > 95%) were from Thermo-Fisher Scientific. The $n^{++}$ Si/$SiO_2$ substrates (15 × 15 $mm^2$ in dimension, dielectric thickness is 230 nm) have prefabricated gold source and drain electrodes (30 nm, with 10 nm ITO as the adhesion layer) on top (channel width = 10 mm, channel length = 2.5, 5, 10, or 20 μm).

### Device and sample fabrication

(i) HMDS treatment: The ultrasonically cleaned substrates were first treated with Oxygen plasma for 3 min, and then they were brought to a HMDS vapor prime oven with a temperature of 110 °C and treatment time of 2 min. (ii) PFS treatment: The cleaned substrates were first treated with UVO for 10 min, and then they were put into a jar filled with PFS vapor at 75 °C and with a treatment time of 1.5 h in a $N_2$-filled glovebox. (iii) Deposition of P3HT and P(NDI2OD-T2) films: The polymers were dissolved in chloroform at 5 mg $mL^{-1}$ and stirred at room temperature for 3 h. The polymer films were deposited by spin-coating at 2000 rpm for 1 min. The samples were then annealed at 80 °C for 30 min. (iv) The whole procedure for the fabrication of PII2T-C4-based FET on OTS-treated substrate with the 1DT- or PFDT-functionalized gold electrodes has been reported in our previous work.[45] Here, we describe the procedure for the deposition of PII2T-C4 films on different silane-treated substrates: The reference substrate with only UVO-treatment (for 10 min) was immediately brought into the glovebox after treatment. The volume of solution for spin-coating was adjusted according to the different hydrophobicity of the substrate surface in order to cover the whole surface. The spin-coating parameters are 2000 rpm and 1 min. All the samples were annealed at 150 °C for 30 min afterwards. The above deposition process of polymer films was all done in the glovebox.

### Electrical and magnetoresistance measurements

The as-fabricated devices of these polymers were connected to the sample holder with copper wires and transferred into the vacuum chamber of Dynacool physical property measurement system. Each device was kept at a high vacuum (< $10^{-5}$ torr) overnight before the measurements. Keithley 2636B semiconductor analyzer was employed to carry out all the electrical measurements. The electrical characteristics were recorded from room temperature down 5 K. The measurement at each temperature was not started until the temperature stabilized for at least 5 min.

To evaluate the MFE, there are, in principle, two methods to determine the resistance values before and after applying $B$. First is to measure the transient resistance by recording the I-V curves (e.g., transfer or output characteristics in the case of an FET) at discrete magnetic fields. Second is to continuously bias the device and monitor the resistance change when scanning $B$. The latter is more commonly used mainly because it can provide a thorough and smooth magnetoresistance line shape. In the context of FET measurements, the latter would also enable a more reliable baseline that excludes the resistance variation induced by the bias-stress effect. $B$ is perpendicular to the substrate, with positive or negative pointing in opposite directions. The scan rate was 1 mT $s^{-1}$ for the ± 100 mT range and 2.5 mT $s^{-1}$ for the ± 500 mT range. Here we describe the measurement of PII2T-C4 device with 500 mT range as an example (see Supplementary Fig. 1), the $I_{DS}$ was monitored by biasing the device with a constant $V_G$ and $V_{DS}$. A 300-s bias was applied before starting the sequence of magnetic field scans from −500 to 500 mT. A built-in script for Keithley was coded to synchronize with the program for the magnetic field scan. In the data analysis stage, the baseline drift due to bias stress was excluded by the stretched exponential function for reliably calculating the magnetoresistance. For the measurements of PII2T-C4 devices with different channel lengths, devices with 2.5 to 20 μm from the same substrate were tested consecutively. The pulsed-gating voltage was applied using a built-in script, and the output pulse parameters, including pulse width and period, were confirmed by an oscilloscope.

**Material characterizations**

*GIWAXS*: The structural characterization was carried out by employing GIWAXS at the Canadian Light Source (CLS) synchrotron on the BXDS-IVU beamline. The polymer films were made with the same recipe on $n^{++}$ $Si/SiO_2$ substrates ($10 \times 10$ $mm^2$ in dimension, dielectric thickness is 90 nm). The beam incidence angle was 0.15 degrees. The images with $4096 \times 4096$ pixels were transformed into the $q$ space and polar space using the software GIDvis.[50]

*Vis-NIR spectroelectrochemistry*: We prepared the samples using similar recipes in device preparation, although the films were prepared in ambient conditions. The Vis-NIR absorbance spectra were acquired in a colorimetry cell (93/G/2, Starna) with a SR-2VN500 visible spectrometer and a Flame-NIR+ spectrometer (Ocean Optics) in a home-made setup controlled by LabVIEW. The incident white light was generated by a halogen light source (HL-2000, Ocean Optics). The polymers were spin-coated on conductive substrates used as working electrodes (2-mm gap between two short-circuited gold electrodes on glass for P3HT and PII2T-C4, and ITO substrate for P(NDI2OD-T2). An Ag/AgCl pellet electrode was used as quasi-reference electrodes (Phymep). KTFSI, $KPF_6$ and $TBAPF_6$ (Merck) were used as salts for aqueous or acetonitrile electrolytes at a concentration of 0.1 mol/L.

*Contact angle measurement*: They were done using OCA-20 from DataPhysics Instrument. The contact angles of pure water and diiodomethane were measured to evaluate the surface energies of $SiO_2$ substrates treated with different silane-SAMs. The substrates used were $n^{++}$ $Si/SiO_2$ substrates ($10 \times 10$ $mm^2$ in dimension, dielectric thickness was ~90 nm).

*ICP-MS*: All the polymer samples, reference samples (JSD-2) and blank samples (Milli-Q water) were measured by a triple quadrupole inductively coupled plasma mass spectrometry (QQQ-ICP-MS, Agilent 8900, Santa Clara, CA). The blank samples were digested and diluted in the same way as the polymer samples to control for background contamination. The samples were digested in tetrafluoroethylene vials with 5 mL of $HNO_3$ (70%) and 1 mL of HCl (36%) using a single microwave digestion platform (Multiwave 5000, Rotor 24 HVT, Graz, AT). The initial cycle consisted of a 10-min ramp-up period to 200 °C, followed by a 30-min heating phase and a final 10-min cooldown. The whole digestate (6 mL) was transferred to a 50-mL polypropylene trace metal-free centrifuge tube (VWR) and the volume was completed to 40 mL with Milli-Q water. Samples were diluted (1/10) with Milli-Q water into trace metal-free tubes before analysis to obtain final $HNO_3$ and HCl concentrations of 2% and 0.25% (v/v), respectively.

**Modeling methods**

*DFT and MD calculations:* The geometries of oligomers of increasing length for P3HT, P(NDI2OD-T2), and PII2T-C4 were optimized at the DFT level using the CAM-B3LYP functional and a 6-31G** basis set. Calculations were performed for neutral, singly charged and doubly charged states. In the latter case, two spin multiplicities were considered to distinguish between polaron pair (triplet) and bipolaron (singlet) configurations. For P3HT and P(NDI2OD-T2), we computed dicationic and dianionic states, respectively. While for the ambipolar PII2T-C4, both dicationic and dianionic states were computed. Atomic charges were subsequently evaluated within the electrostatic potential framework. We regarded one unit corresponds to one thiophene in P3HT, while in P(NDI2OD-T2) and PII2T-C4, one monomer consists of three and four units (including two units linked by the C=C bond in the isoindigo part and two thiophenes), respectively. Spatial localization of the doubly charged states (Fig. 4b) was determined from the difference between the electronic densities of the charged species and those of the corresponding neutral systems.

To quantify the extent of π-conjugation along polymer backbones in the amorphous phase, atomistic MD simulations were performed following the reported protocol[51]. Initial configurations were generated by randomly placing 24 oligomers within a large simulation box (~300 Å × 300 Å × 300 Å), with oligomer lengths corresponding to n = 10 for PII2T-C4, n = 12 for P(NDI2OD-T2), and n = 45 for P3HT. The systems were first equilibrated under low-density conditions using a 500-ps NVT simulation at 1000 K to ensure random spatial distribution and minimize initial structural bias. This was followed by a stepwise compression and cooling procedure consisting of five successive 500-ps NPT simulations at 1 atm, with temperatures decreasing from 1000 K to 298 K (1000 K, 500 K, 400 K, 350 K, and 298 K). Subsequently, long NPT simulations (tens of nanoseconds at 298 K and 1 atm) were performed until convergence of the system density was achieved. Production trajectories were then generated over 10 ns, during which configurations were recorded every 400 ps for structural analysis (501 snapshots).

All simulations were carried out using the Materials Studio package. The reported force fields for P3HT[52] and P(NDI2OD-T2)[51] were used. For PII2T-C4, the Dreiding force field was re-parameterized as described in Ref.[51]. In particular, torsional potentials governing rotations between adjacent conjugated units, between conjugated backbones and alkyl side chains, and along the side chains were refined and benchmarked against MP2/cc-pVTZ calculations. Atomic charges for the conjugated cores were obtained by fitting the electrostatic potential calculated at the MP2/6-31G** level on isolated dimers.

Torsional angles between successive conjugated units were measured from 501 snapshots of the equilibrated MD trajectories. To account for the equivalence of *cis* and *trans* planar conformations, each torsion angle θ (defined in the range [0°, 360°]) was mapped onto a minimal deviation from planarity: $\delta(\theta) = \min(|\theta|, |\theta-180°|, |\theta-360°|)$ such that $\delta \in [0°, 90°]$. We considered that two subsequent units are conjugated when $\delta(\theta) \leq \theta_c$, where $\theta_c$ is a predefined cutoff angle. Three values (15°, 30°, and 45°) were used to assess the dependence of the conjugation length on the predefined cutoff angle.

For each polymer chain, conjugated segments were identified by scanning the sequence of consecutive torsions fulfilling the above criterion. Segments of length of one, corresponding to isolated monomers without adjacent planar connections, were excluded from the statistical analysis, as they do not contribute to effective conjugation. All segments identified across all chains and all snapshots were aggregated to construct a probability distribution, describing the statistical distribution of conjugation lengths in the amorphous phase (Supplementary Fig. 21). In addition, for each chain and each snapshot, the maximum segment length $L_{max}$ was extracted (Fig.

4c–e), yielding a dataset of size $N_{chains} \times N_{snapshots}$. This quantity provides a measure of the longest locally conjugated segment for a given polymer conformation.

*Numerical modeling of bipolaron formation at the interface*: detailed in Supplementary Section 1.2.

*DFT calculation on protonation of polymer chains*: Calculations of protonation affinities for P3HT and PII2T-C4 were performed using the ωB97X-D functional[53] with the 6-31+g(d,p) basis set. Geometry optimizations for PII2T-C4 were carried out on a pristine model consisting of two monomers ($(II\text{-}2T)_2$), with side chains replaced by methyl groups. Protonation affinities were then obtained from the energy difference between the optimized pristine structure and the optimized structure of the same dimer model protonated at each possible site, computed at the same level of theory. The various protonation sites are illustrated in Supplementary Fig. 23; the same procedure was applied to P3HT using a nine-unit model ($T_9$) with methyl groups representing the side chains. When protonation occurs at key sites that connect conjugated cycles, significant geometric changes arise from the loss of conjugation. These sites are also where conjugation-breaking defects are expected to be naturally present in the polymer. Therefore, we additionally calculated corrected protonation affinities by taking the defect-containing structures (before protonation) as the initial conformation instead of the pristine structure. Specifically, we removed the proton from those significantly perturbed geometries and computed the energy of the resulting deprotonated structures without performing further geometry optimization, which we considered as the initial conformation in the calculation of corrected proton affinities (Supplementary Table 7).

## Acknowledgements

This work was carried out within the framework of the IdEx University of Strasbourg, thanks to the MAESTRO grant. O.B. acknowledges the French national research agency (ANR) for funding the ANR CAROT (ANR-24-CE06-7350-01).

We thank Ingo Salzmann of Concordia University, as well as Beatriz Moreno and Chang-Yong Kim at the CLS-IVU beamline for the assistance on the setup of GIWAXS experiments. We acknowledge Marta Gabriele and Matthieu Moingt of UQAM for the ICP-MS measurement.

## Author contributions

E.O. and Z.Y. conceived this study. E.O. supervised the project. Z.Y. fabricated the samples and performed the electrical and magnetotransport measurements as well as all the film characterizations. V.L. and Y.O. carried out the DFT and MD simulations. M.B.R. ran the modeling of interfacial bipolaron and the DFT calculation of protonation effect. O.B. performed the spectro-electrochemical characterizations.